# High Uniformity GaN Micro-pyramids and Platelets by Selective Area Growth

Changhao Li[1], Vitaly Z. Zubialevich[1], Peter J. Parbrook[1, 2], Brian Corbett[1], and Zhi Li[1*]

[1] Tyndall National Institute, University College Cork, Lee Maltings, T12 R5CP, Cork, Ireland

[2] School of Engineering, University College Cork, College Road, T12 HW58, Cork, Ireland

*Email: zhi.li@tyndall.ie

**ABSTRACT:** The development of uniform GaN micro-pyramids and platelets via selective area growth is a critical step toward advancing III-nitride device technologies, particularly for micro-light-emitting diode applications. This work investigates the origins of morphological non-uniformity in micro-pyramids and micro-platelets grown by metal-organic chemical vapor deposition (MOCVD). We observe that a direct one-step growth approach leads to significant growth rate inhomogeneity across arrays. To shed light on this issue, we examine the mechanisms driving non-uniformity and explore process modifications aimed at mitigating these effects. Building on these insights, we propose a controlled multi-step growth strategy that combines sequential growth and thermal treatment phases. This approach is demonstrated to enhance surface morphology and structural regularity. The work contributes to the broader objective of enabling scalable, high-precision GaN microstructure fabrication for next-generation optoelectronic applications.

## INTRODUCTION

III-nitrides have been widely used in various electronic and optoelectronic applications from light-emitting diodes (LEDs), laser diodes (LD), transistors and photodetectors due to their wide bandgaps, high electron mobilities and exceptional chemical stability[1–3]. Among these, micro-light-emitting diodes (micro-LEDs) have emerged recently as a promising candidate for next-generation display technology owing to advantages such as high brightness, high resolution and long lifetime[4]. However, the efficiency of these devices, especially for red-emitting devices and when the chip sizes less than 10 μm, is far from satisfactory mainly due to the non-radiative surface recombination[5–7]. Epitaxially-grown microscale GaN structures, including pyramids, platelets and nanowires, offer a unique solution to overcome the efficiency challenges owing to their core–shell geometry and strain-relieving properties[8–13]. Compared to conventional devices based on planar substrates, these three-dimensional microstructures can not only reduce the lattice mismatch induced strain and the quantum-confined Stark effect (QCSE)[14,15], but also improve the light extraction by assisting the escape of the photons[16]. In addition, the bottom-up approach can avoid the side wall damage[17], a common issue facing the conventional device fabrication process due to the dry etch steps required.

The formation of micro-pyramids and platelets is typically conducted by the selective area growth (SAG) technique through metal-organic chemical vapor deposition (MOCVD)[18] and/or hydride vapor-phase epitaxy (HVPE)[19]. However, the shape non-uniformity of these microstructures, particularly at small sizes of a few micrometres or less, remains a significant challenge during the epitaxial growth process, hindering their practical applications. The difference in growth rate between various facets is affected by growth temperature, carrier gas ratio, etc. and is what defines the ultimate shape of microstructures. Unfortunately, when targeting micro-platelets/pyramids, these growth processes have issues that ultimately lead to irregular cross-section, truncated or plagued with V-pits tops, height-inconsistent across arrays, etc.[20–22]. Previously reported efforts to suppress such non-uniformities have included optimizing temperature, adjusting ammonia partial pressure, utilizing various carrier gas ambients and inserting a seed layer[23–25]. In this work, we present a comprehensive study on the SAG of GaN micro-pyramids and platelets with the aim to achieve their high uniformity. We investigate the effect of initial GaN template, growth parameters, and post-growth annealing. We first investigate the dependence of the resulting microstructure morphologies on growth temperature and $H_2$ partial pressure for different types of GaN templates. Based on our observations, we conclude on the undermining mechanism, proposing and successfully realizing an approach capable of overcoming complications of a conventional growth.

## EXPERIMENTAL

Three types of templates, prepared on sapphire substrate were used: Type *A* considered of a 2 µm thick *n*-GaN/*u*-GaN substrate grown on a commercially provided 25 nm thick sputtered AlN wafer (Kyma Tech, USA). Type *B* was a commercially provided 2 µm *n*-GaN/*u*-GaN structure grown by MOCVD (Enkris, China), and finally Type *C* was simply 25 nm commercial sputtered AlN (Kyma Tech, USA) on sapphire. The crystalline quality of the three templates was evaluated by X-ray diffraction rocking curve measurements of the symmetric (002) reflection. The corresponding full width at half maximum (FWHM) values are 83.4 arcsec for Type A, 264 arcsec for Type B, and 17 arcsec for Type C, respectively, indicating significantly different crystal qualities among the templates. Atomic force microscopy (AFM) was also performed to assess the surface morphology of the templates; representative AFM images are provided in the Supplementary Information (Supplementary Figure S1).

The growth of initial GaN-templates (type *A*) as well as GaN micro-pyramids and platelets was realized in an Aixron 3 × 2″ close-coupled showerhead-type MOCVD reactor. Trimethylgallium (TMGa) and ammonia were used as sources of Ga and N, respectively. Growth reactor pressure was fixed at 150 mbar, while temperature was fixed at 1060°C for templates and varied from 825°C to 970°C for the SAG of micro-pyramids. The temperature reported here is believed to be the wafer temperature, measured with a pyrometer of an in-situ monitoring tool (LayTec EpiCurveTT). The carrier gas ratio was controlled by adjusting the individual $H_2$ and $N_2$ gas flows keeping the total flow at 4000 sccm. The thermal annealing process was conducted at 300 mbar under an $N_2$+$NH_3$ ambient, with an ammonia flow rate comprising ~1/3 of the total flow (60 mmol/min) for 15 min.

To prepare the templates for the selective-area growth, a 120 nm-thick $SiN_x$ layer was first deposited on the templates using an inductively coupled plasma chemical vapor deposition system (SENTECH). Circular holes with a diameter of 2 µm arranged in triangular array with a pitch of 3 µm were then patterned by standard optical lithography. Additional experiments examining the influence of opening diameter and array pitch on the morphology of GaN pyramids are provided in the Supporting Information (Figure S3). The $SiN_x$ mask was dry etched using a mixture of $SF_6$ and $C_4F_8$ gases in an inductively coupled plasma etcher (SPTS Synapse) to transfer the lithographic pattern (Figure 1). The potential organic residues were then removed using piranha solution. A buffered oxide etch (BOE) was used to partially etch the $SiN_x$ layer, ensuring the complete opening of the microholes. A 45wt% potassium hydroxide solution (KOH) at 90 °C was used to remove damage caused by the dry etch process before loading into the MOCVD growth chamber.

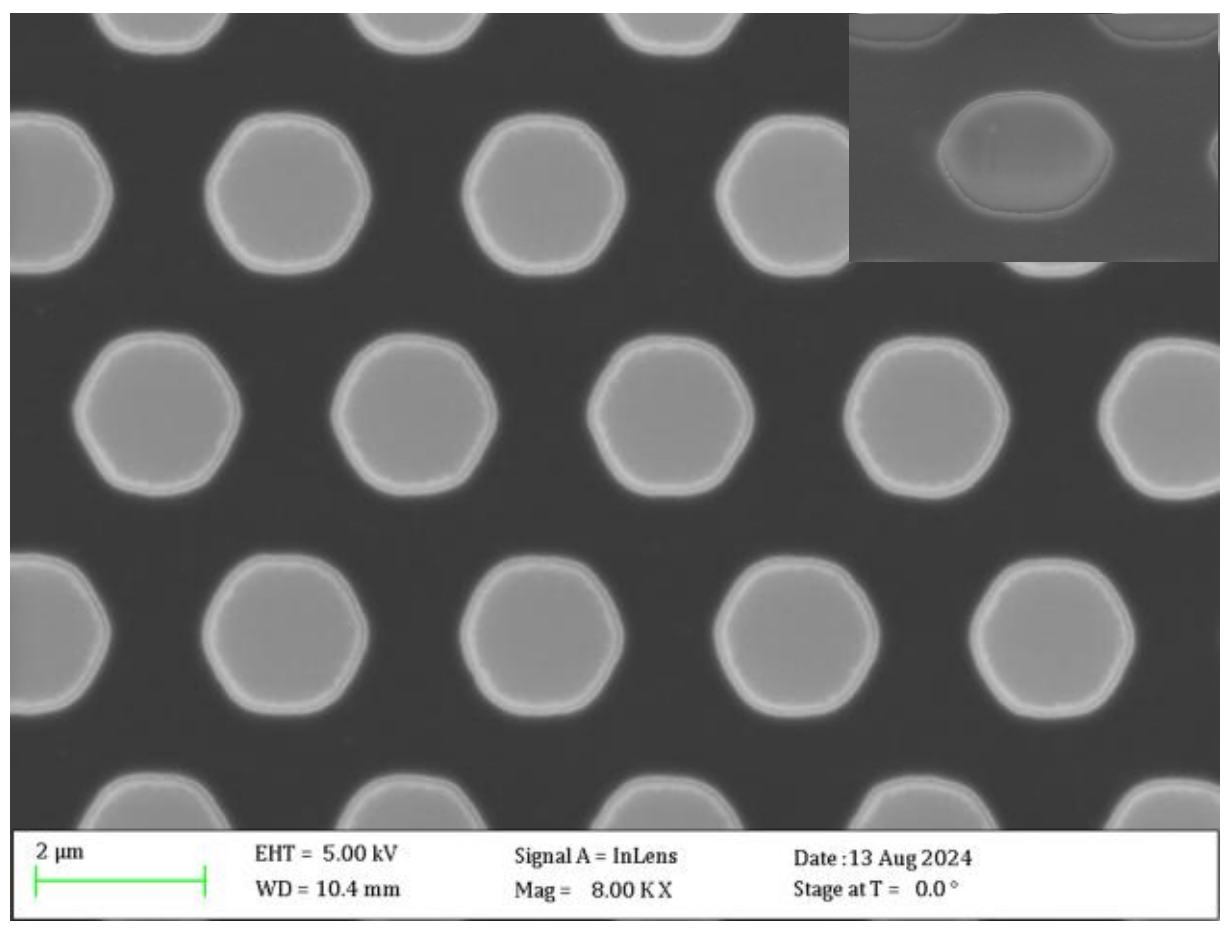

Figure 1. Top view (main figure) and 45°-tilted view (inset) of the patterned SAG GaN template. Lighter regions correspond to the exposed GaN surface.

The morphology of GaN micro-pyramids and platelets prior and after growth was assessed using a Carl Zeiss AG - SUPRA 40 scanning electron microscope (SEM) with an electron energy of 5 keV and a working distance of approximately 10 mm. The surface morphology of the top of the platelets and the heights of the platelets were investigated by a Bruker Dimension Icon atomic force microscope (AFM) using a tapping mode.

## RESULTS AND DISCUSSION

**Effect of Growth Temperature on GaN Platelets Growth.**

To make an initial understanding of the growth of micro-pyramids/platelets, a series of samples were prepared using varied temperature growth (825°C, 855°C, 890°C and 970°C) on type *B* templates in a hydrogen ambient. Temperature has been previously reported to be a critical parameter that strongly affects the GaN morphology[26], and our data confirms that. The resulting morphology with growth temperature are shown in Figure 2, *a*-*d*. At the lowest temperature of 825 °C, while reasonable height/size uniformity was obtained, the top *c*-plane facets of the platelets were severely plagued with V-pits, which results in corrupted/distorted apexes when platelets are grown into completed pyramids are shown in Figure 8, *b*. With the increase of temperature to 855°C, platelets with much smoother surfaces and well-developed hexagonal cross-section were formed. However, a small fraction of platelets appeared to be growing faster than the rest, and this subgroup were found to have residual V-pits on their top facets. This effect is also observed at 890°C with a decreasing density of V-pits for the taller subgroup of truncated pyramids. Finally, at the highest applied temperature of 970°C (Figure 2, *d*), the size non-uniformity is extreme, with many openings presenting no growth at all. Those rare GaN microstructures that managed to grow while demonstrating rather good surface morphology (without any signs of V-pits), were characterized by a high inhomogeneity in their size and height.

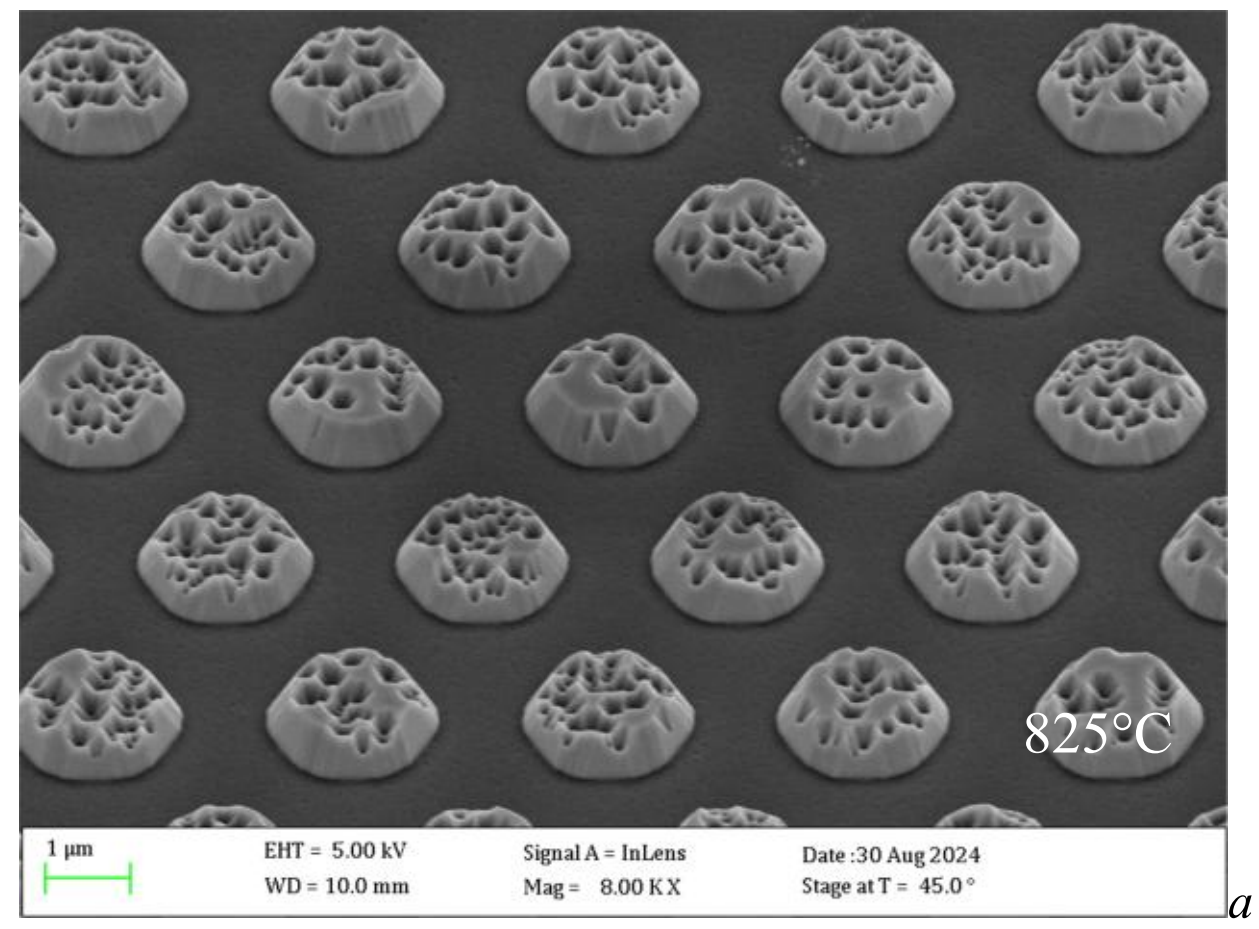

*a*

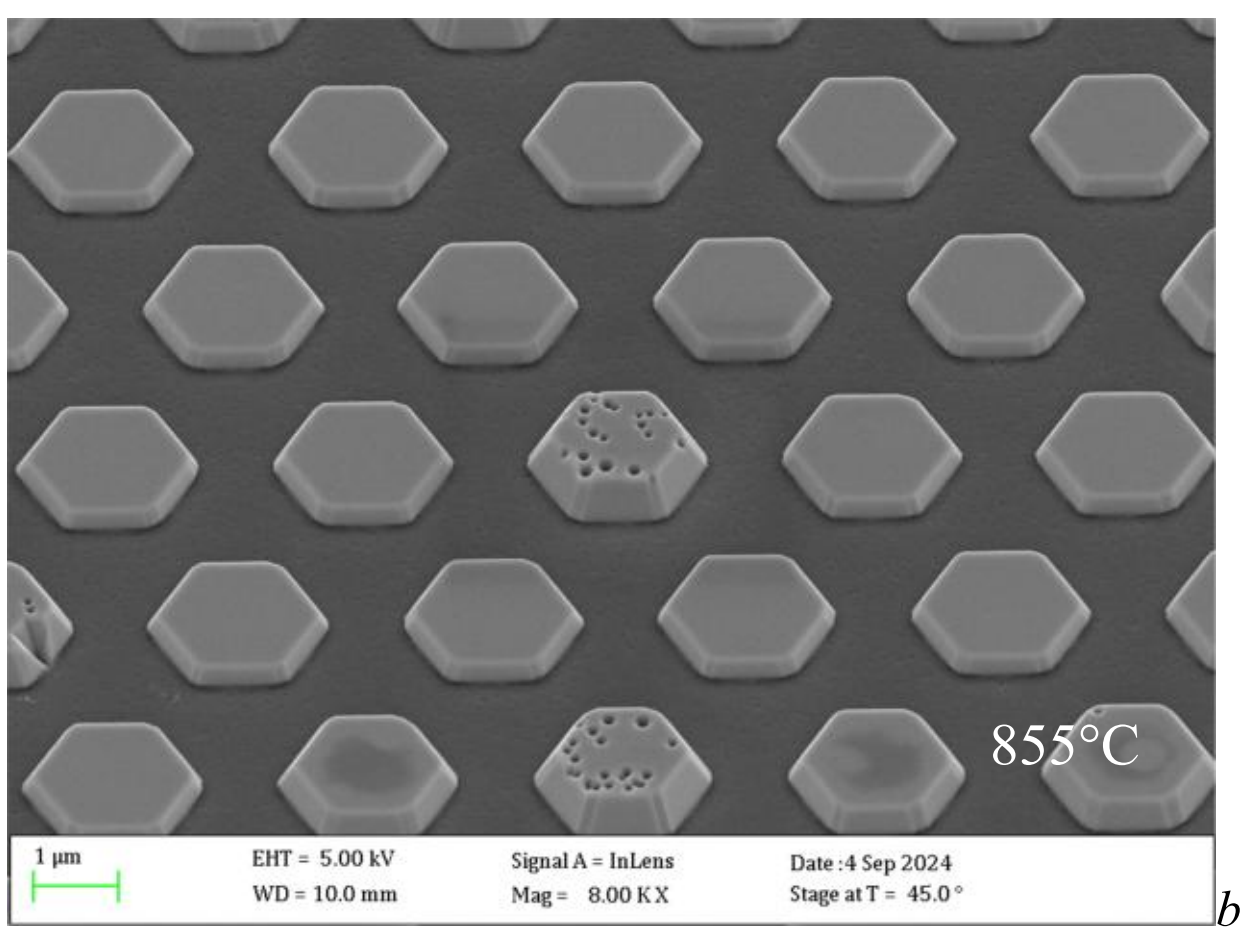

*b*

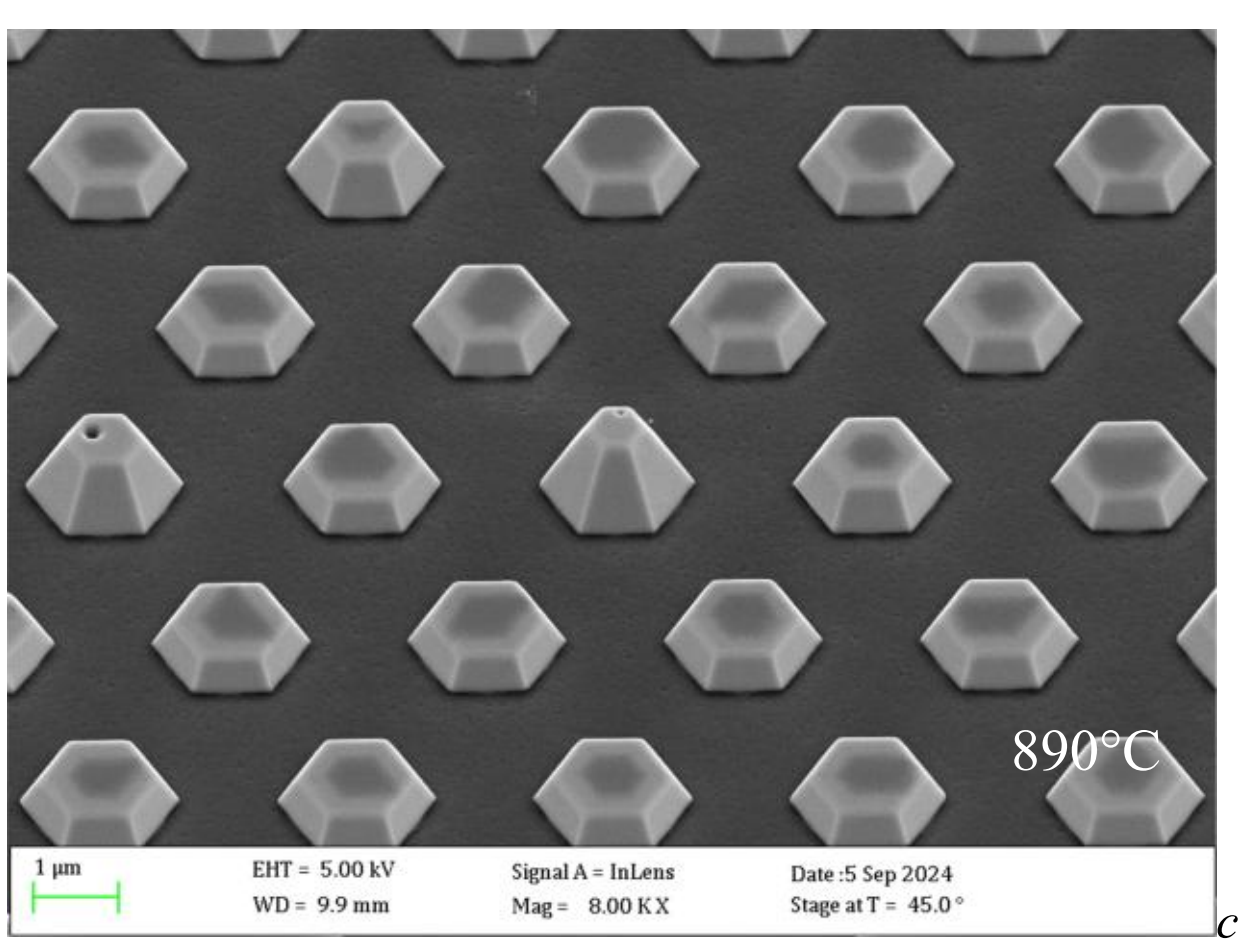

*c*

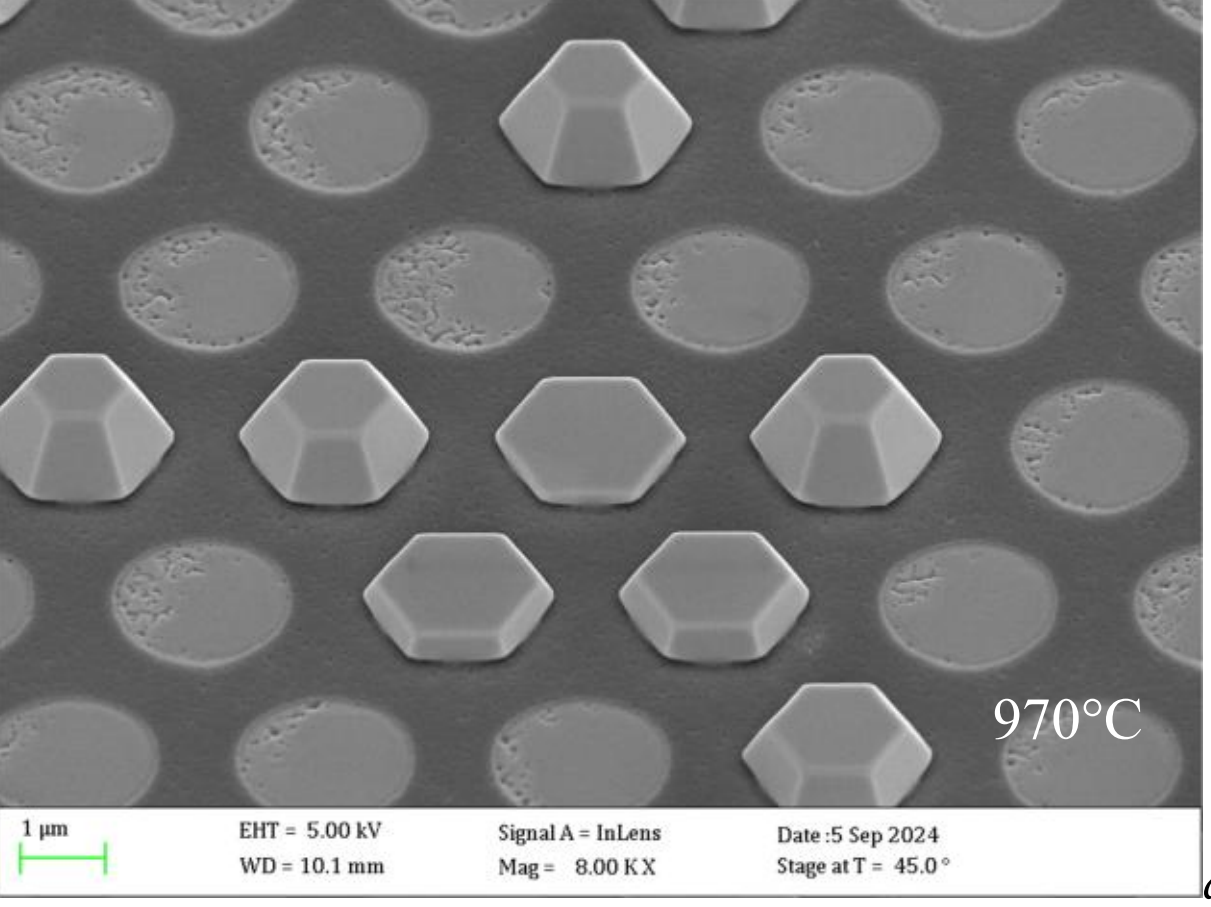

*d*

Figure 2. GaN platelets grown in $H_2$-rich ambient at different temperatures: (*a*) 825°C, (*b*) 855°C, (*c*) 890°C, (*d*) 970°C.

In general, it is clear that mutual competition between individual microstructures is only possible when the effective Ga adatom diffusion length is greater than the array pitch. From our observations, we conclude that this is not the case at the lowest growth temperature of 825ºC. At this temperature, the diffusion distance of Ga adatoms is such that they can only migrate to a closest microstructure, resulting in all GaN platelets tending to grow with similar growth rates as they cannot “steal” Ga adatoms from each other. Unfortunately, microstructures cannot provide a smooth surface morphology of their *c*-plane facets under these conditions.

It is worth noting the absence of V-pits on the *c*-plane facets of short platelets in Figure 2, *b* while they are present consistently in tall ones. As we previously reported for AlGaN microhoneycomb structures[27], the evolution of V-pits is governed by the ratio between the growth rates along the *c*-plane ($GR_{c\text{-plane}}$) and the adjacent semipolar facets ($GR_{s.p.}$). The threshold value of this ratio ($r_{GR,\ thr} = GR_{s.p.}/GR_{c\text{-plane}}$ which separates the regimes of V-pit closure versus its expansion) depends on the particular crystallographic orientation of the semipolar facets involved (through the angle between them and the *c*-plane), while the growth rates themselves typically depend on growth parameters i.e. temperature, pressure, V/III ratio, etc.

As shown in Figure 8, *b*, top-view SEM images of our low-temperature-grown microstructures clearly indicate that the V-pits are aligned in the same orientation as the pyramids/platelets themselves, confirming that they belong to the $\{1\bar{1}01\}$ family of planes. From the observed V-pit formation, we infer that, at these lower temperatures (825°C), the growth rate along the $\{1\bar{1}01\}$ directions $GR_{\{1\bar{1}01\}}$ is very slow and specifically, below at least one-half of the growth rate along the *c*-plane $GR_{c\text{-plane}}$. This conclusion is supported by geometric considerations[27]: the cosine of the angle (~62°) between the *c*-plane and $\{1\bar{1}01\}$ facets represents the critical (threshold) ratio ($r_{GR,\ thr} \approx 0.47$) for maintaining a *c*-plane surface area constant in the presence of V-pits. For $r_{GR}$ smaller than 0.47, V-pits are expected to grow and for $r_{GR}$ larger than 0.47, they will shrink and disappear (or not form if they were not present initially). However, the initiation of new V-pits from a nominally smooth *c*-plane surface may require an even lower semipolar-to-*c*-plane growth rate ratio.

Upon increasing the growth temperature to 855°C and to some extent even to 890°C, the *c*-plane surfaces of most platelets (notably mostly short ones!) become smooth (Figure 2, *b*, *c*), which means that the ratio $r_{GR}$ increases with temperature. The fact that the short platelets have no V-pits is consistent with the proposed mechanism. As *c*-plane growth rate of short platelets is obviously slower than that of tall ones, the growth rate ratio $r_{GR}$ for them is also proportionally larger, and apparently larger than $r_{GR,\ thr}$ thus preventing V-pits formation, despite tall ones still having them. This can explain the observed difference in *c*-plane surface morphology of different micro-platelets despite identical thermodynamic parameters at which they are grown.

While the above considerations account for the morphological differences of the top *c*-plane facets, they do not explain why the growth rate along the *c*-direction differs between micro-platelets. To clarify this point, a series of AFM measurements was conducted, as described in the following section.

**Mechanism of GaN platelet non-uniformity**

As seen in Figure 2, *d*, high growth temperatures are highly effective in suppressing V-pit formation. However, under these conditions, most of the openings appear to struggle to initiate growth. To address this issue, we implemented a two-step growth approach: a low-temperature "nucleation" phase followed by a medium-high temperature growth phase, using 820 °C and 950 °C for the respective steps. Figure 3, *a*, shows a representative SEM image of the resulting GaN SAG. While this strategy effectively mitigates the formation of V-pits, it does not resolve microstructure-to-microstructure height non-uniformity. Two distinct populations of platelets are observed: one taller and one shorter, which are distributed randomly across the array, with the shorter platelets being the majority.

The absence of V-pits on the *c*-plane facets allowed for high-resolution atomic force microscopy (AFM) scans of individual micro-platelets. We performed $1 \times 1\ \mu m^2$ AFM scans on both short and tall micro-platelets, representative examples of which are shown in Figure 3, *b* and *c*, respectively. The resulting AFM maps clearly reveal the origin of the observed growth rate disparity.

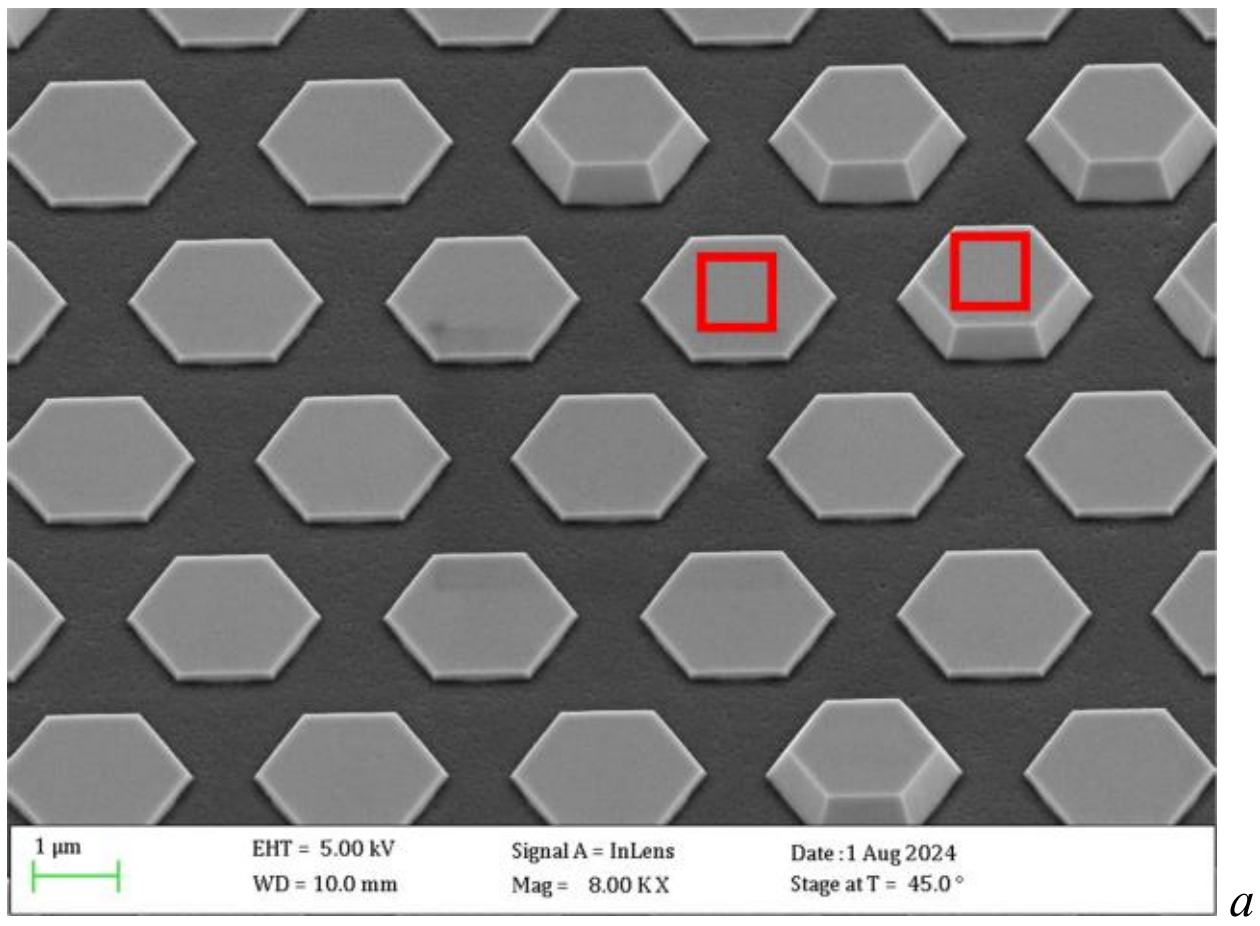

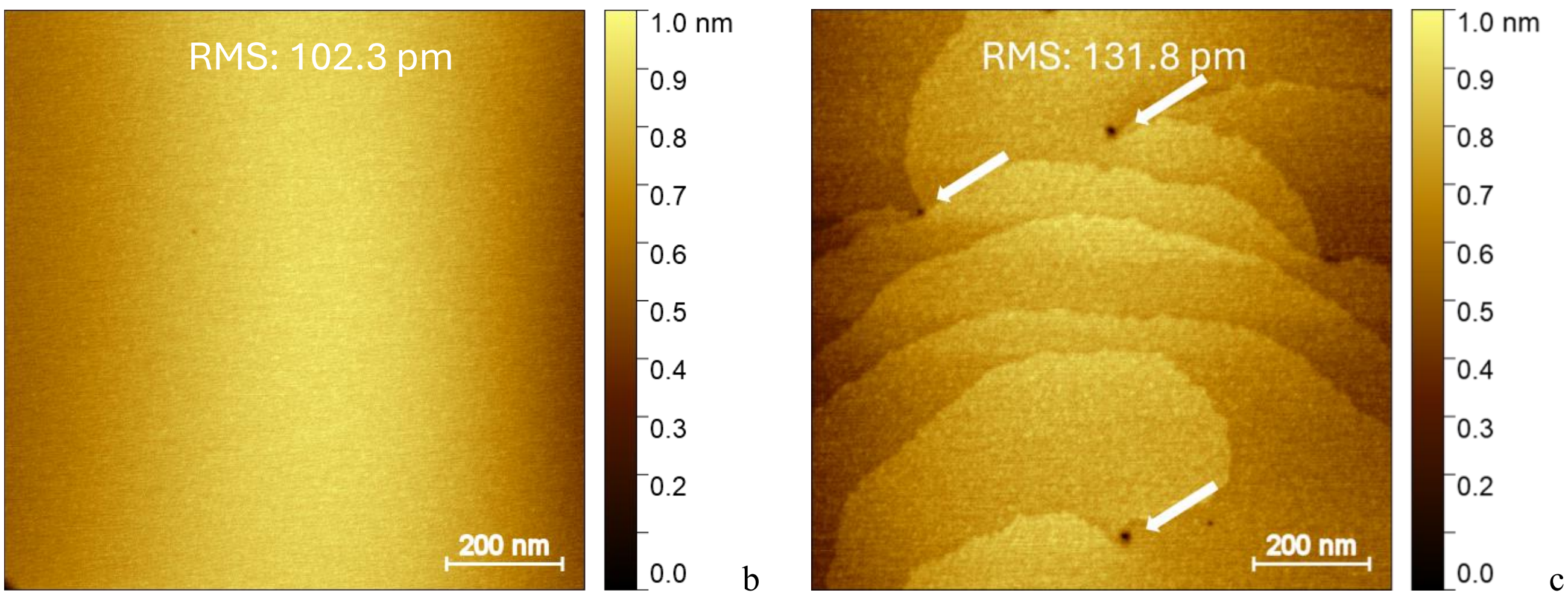

Figure 3. SEM (45°-tilted view) of micro-platelets grown using two temperature approach (*a*) and typical surface morphology of top *c*-plane facets as measured by AFM for short (*b*) and tall (*c*) platelets.

As seen from Figure 3, *b*, the shorter platelets exhibit an atomically smooth surface, with a root-mean-square (RMS) roughness of only 0.10 nm and no observable atomic steps across the whole scanned area. In contrast, *c*-plane facets of the taller platelets, while only marginally rougher in terms of RMS (0.13 nm), exhibit multiple atomic steps and prominent dark spots. These dark features correspond to the cores of dislocations of which at least 3 (in the example presented in Figure 3, *c*) correspond to screw- or mixed-type ones.

The presence of dislocations with a screw component of the Burgers vector within the microhole significantly enhances the local growth rate. The underlying mechanism is schematically presented in Figure 4. Screw dislocations act as continuous sources of atomic steps, which serve as energetically favorable incorporation sites for Ga adatoms, enabling spiral growth around the dislocation core[28,29]. Consequently, *c*-plane facets that host screw dislocations function as efficient sinks for incoming adatoms, enabling accelerated vertical growth. In contrast, micro-platelets that are free of screw/mixed-type dislocations, once a monolayer is completed, lack available step edges for continued incorporation. Ga adatoms on these atomically smooth surfaces must diffuse extensively before a new nucleation step can form. Often, these adatoms are ultimately captured by neighboring microstructures containing screw dislocations, driven by the local gradient in adatom concentration. As a result, micro-platelets lacking screw dislocations grow significantly more slowly than those with one or more such defects.

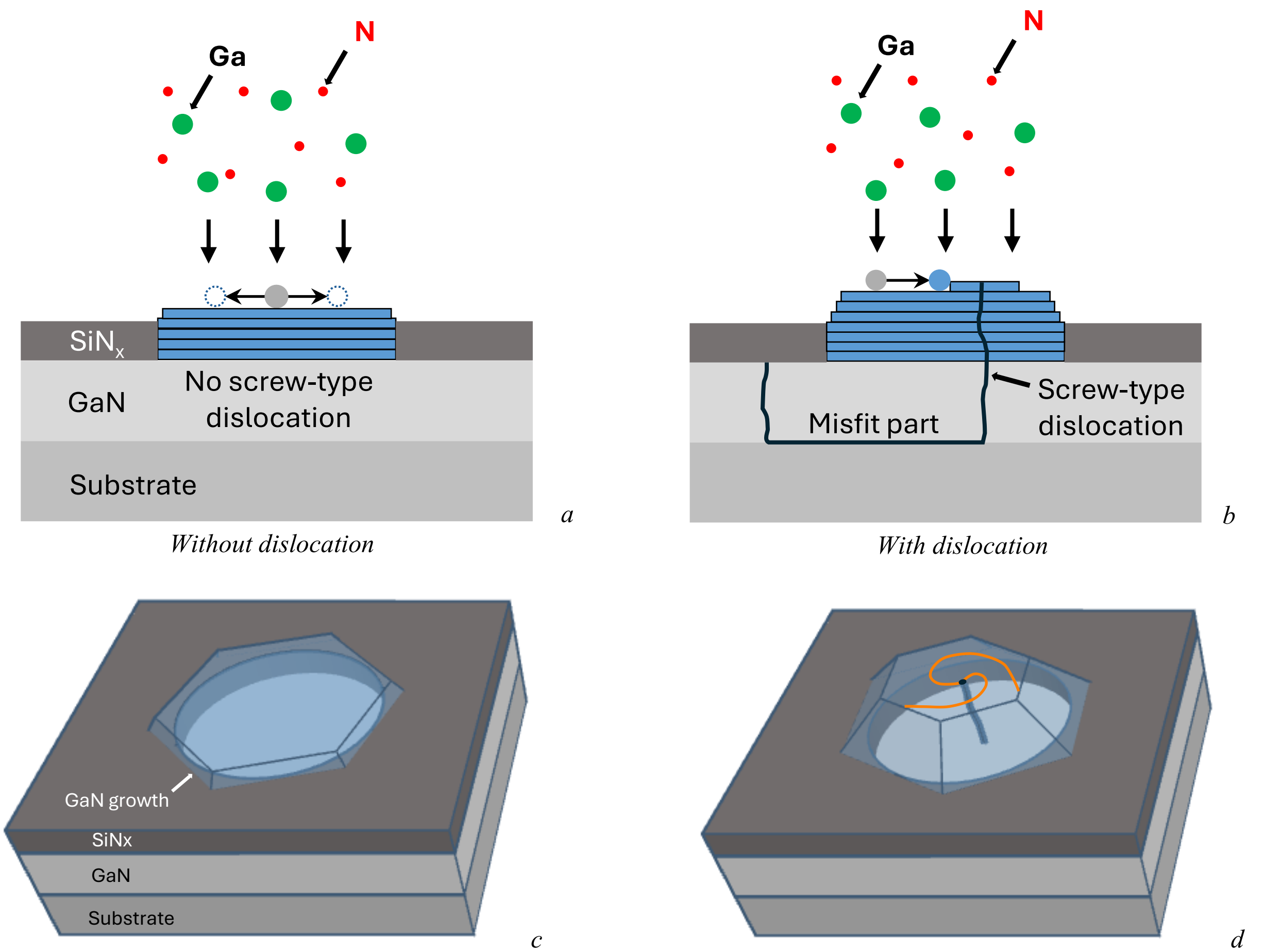

**Description:** grey circles represent just arrived Ga adatoms, open dotted circles are Ga-adatoms still not incorporated to the lattice after extended surface diffusion, blue circles are representing Ga adatoms that found an atomic step and incorporated into the lattice there, green and red circles represent gallium and nitrogen atoms in the gas phase (disregarding actual precursor cracking).

*Figure 4.* Schematic illustration of screw dislocation enhanced vertical growth rate of micro-platelets: *a*) case of dislocation-free microstructure, *b*) case of a microstructure with a screw dislocation, *c*) and *d*) morphology of the resulting microstructures at high temperature when no V-pits are formed for both short and tall platelets.

This phenomenon is consistent with some prior studies. For instance, Akasaka *et al.*[30,31] associated the increased height of homoepitaxially grown GaN hexagons with the presence of screw-type dislocations. More recently, Cai *et al.*[32] reported a correlation between the density of tall InGaN submicron platelets and the density of screw dislocations in the underlying GaN templates used for selective area growth.

This mechanism also provides a consistent explanation for the temperature-dependent trends observed in Figure 2. At sufficiently low temperatures, nucleation of a new monolayer is relatively easy due to the enhanced "stickiness" of Ga adatoms. Under these conditions, the growth rate becomes less dependent on the presence of screw dislocations or atomic steps introduced by the associated spiral growth. As a result, the microstructures exhibit relatively uniform vertical growth rates, as shown in (Figure 2, *a*). The observation that microstructures segregate into two distinct height groups suggests that having more than one screw dislocation within a *c*-plane facet does not confer additional growth rate advantages. This implies that even a single dislocation induced atomic step per microstructure is sufficient to incorporate the full flux of Ga adatoms, and that multiple steps must compete for the same adatom population, at least until the temperature increases to the point where long-range adatom diffusion from neighboring structures becomes a significant factor. This may explain the greater height variation observed among the taller microstructures in Figure 2, *d*. Additionally, some natural height variation among the tall structures can be attributed to the spatial location of the screw dislocation core within the *c*-plane facet. When the dislocation core is located within ~50–70 nm of the facet edge, image forces become non-negligible[33,34], potentially pulling the dislocation toward the semipolar facet. Once the dislocation terminates at the semipolar sidewall, growth of that microstructure may proceed similarly to that of dislocation-free (short) ones.

**Effect of Template on GaN Platelets Growth.**

To further validate our hypothesis regarding the role of screw dislocations in determining microstructure growth rates, we investigated selective area growth on the three templates which have markedly different screw dislocation densities: in-house grown GaN on a commercially available (Kyma Tech, USA) AlN/sapphire (Type *A*); commercial GaN/sapphire (Enriks, China) (Type *B*); and Kyma's AlN/sapphire (Type *C*). Type *B* templates consist of thick GaN-on-sapphire with screw dislocation densities representative of typical commercial heteroepitaxial GaN (>$10^8$ $cm^{-2}$). The Type *C* template, by contrast, comprises thin AlN-on-sapphire and is characterized by extremely low densities of screw-type dislocations albeit with a higher density of edge-type ones. The Type *A* template results, which were used for the study above (Figure 2), was grown on the latter and also is characterized by rather low screw dislocation densities (low $10^7$ $cm^{-2}$).

The results from the comparative study strongly support our proposed growth mechanism. For the type *B* sample, a majority of microstructures grown at 855°C appears as tall platelets (Figure 5, *a*), while there is a clear subset of lower platelets. In the sample *A* grown at the same conditions, as we remember, there was also height variability but with tall subpopulation being minority (Figure 2, *b*). Expectedly, the type *C* sample with virtually no screw dislocations shows no such height variability: its microstructures are all effectively the same size, with no distinctly "tall" or "short" subpopulations (Figure 5, *b*).

It is again worth noting that the short platelet subpopulation in the type *B* templates exhibits no observable V-pits on their top *c*-plane facets, similar to what was found in type *A*. The fact that the *c*-plane facets of all platelets in type *C* are covered with V-pits simply indicates that, under these growth conditions, already the intrinsic (i.e., not additionally reduced by dislocation-induced *c*-plane growth acceleration) growth-rate ratio $r_{\mathrm{GR}}$ is below the threshold value $r_{\mathrm{GR,\,thr}}$. We emphasize that the V-pits observed here are not necessarily associated with dislocation cores but are more likely spontaneously nucleated features. The apparent correlation between V-pits and screw dislocations in samples *A* and *B* arises from the locally increased *c*-plane growth rate induced by the dislocations, rather than from the dislocations themselves serving as nucleation sites.

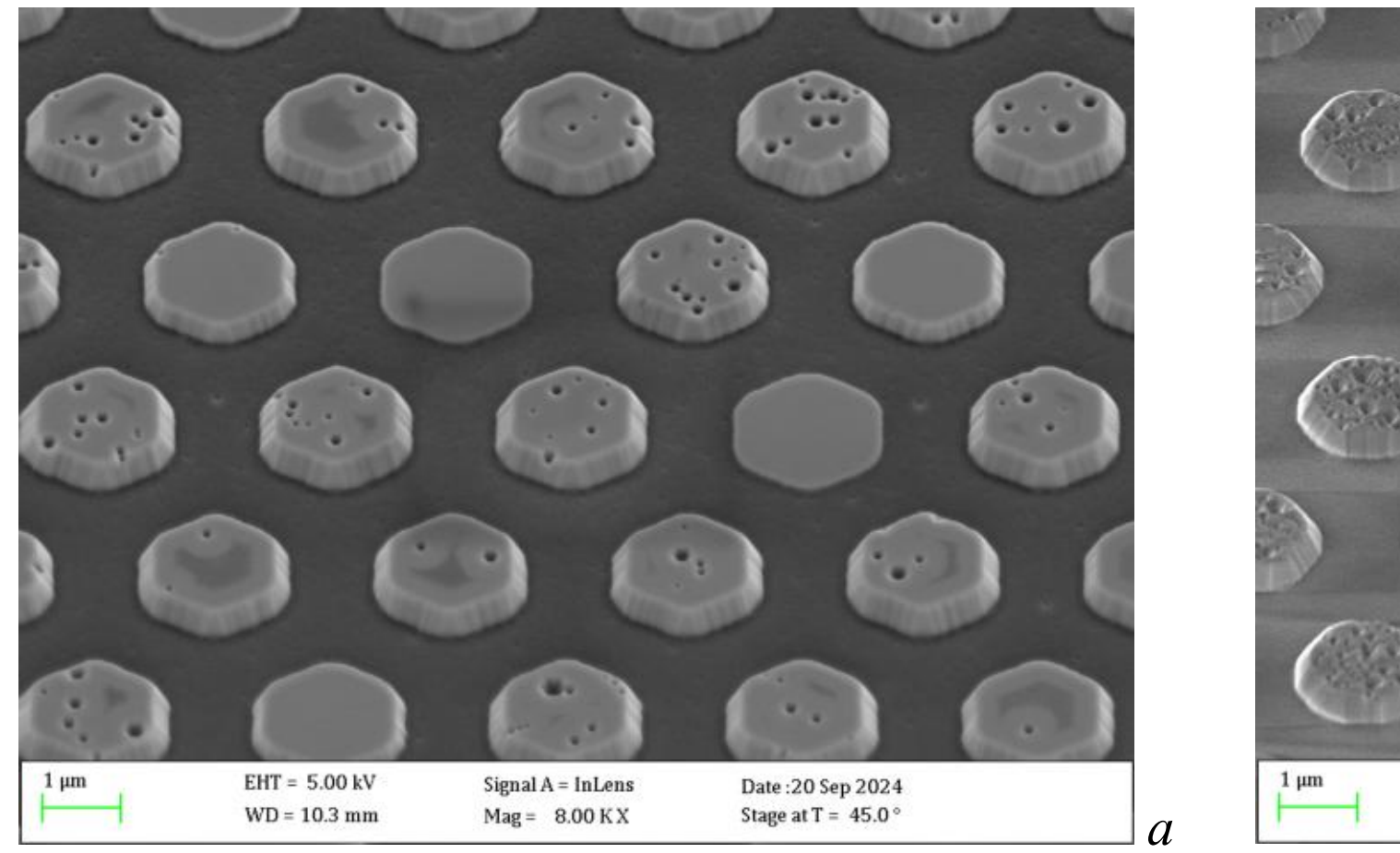

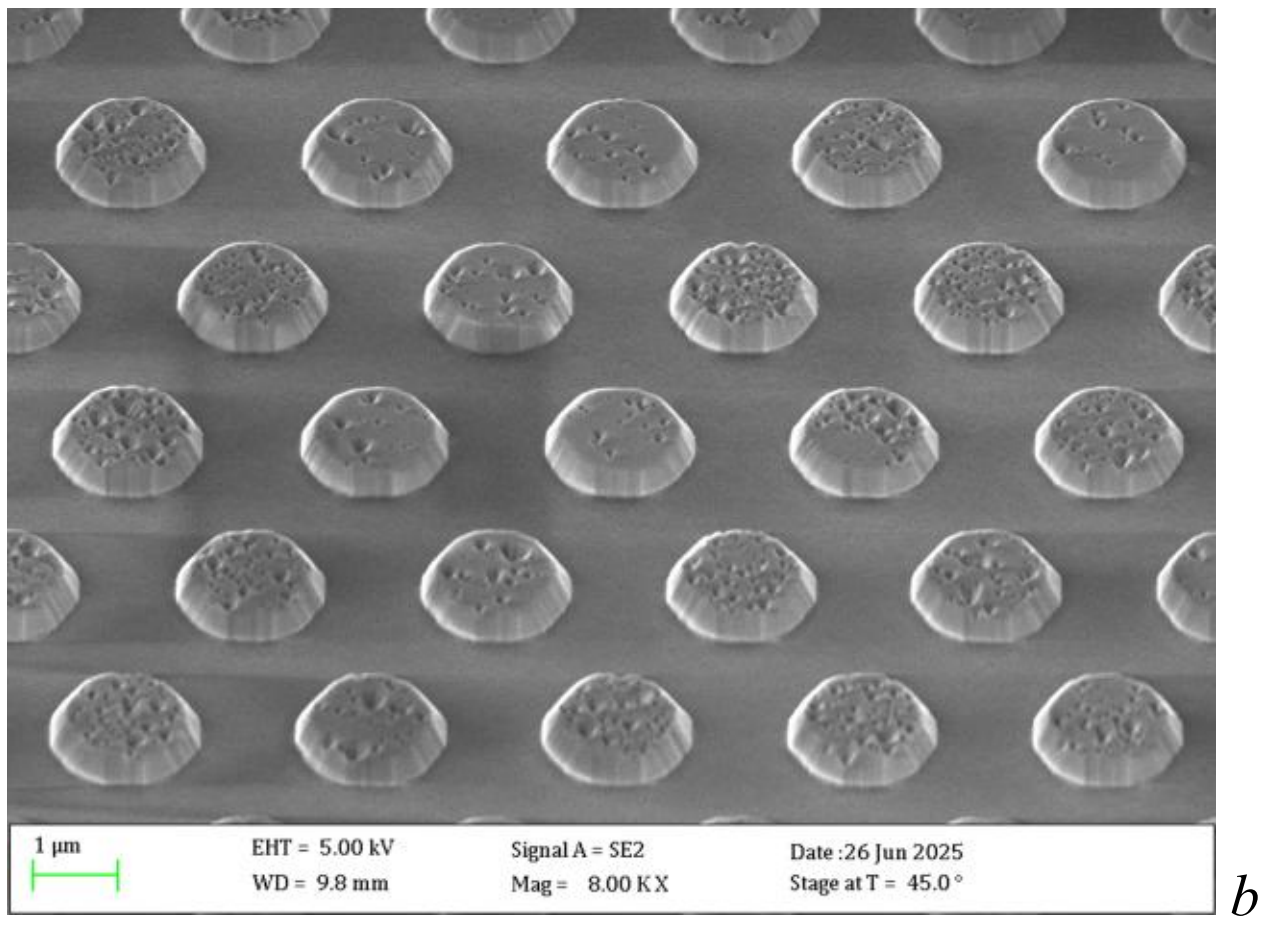

*Figure 5.* Typical surface morphology of platelets on different templates: (*a*) template *B*, (*b*) template *C*, grown at the same conditions as template *A* in Figure 2, *b*.

To support this qualitative observation, we carried out a more quantitative analysis. Screw dislocation densities ($D_{\mathrm{screw}}$) were estimated from the full width at half maximum (FWHM) of the (002) X-ray rocking curve, using the established formula[35]: (1):

$$D_{\mathrm{screw}} = \frac{FWHM_{(002)}^2}{2 \cdot \ln(2) \cdot b_{\mathrm{screw}}^2}, \qquad (1)$$

where $b_{\mathrm{screw}}$ is the Burgers vector of GaN in the [0001] direction (0.5189 nm). Based on these values, we calculated the approximate number of screw dislocations exposed to the surface in specific regions of our array.

Given a microhole diameter of ~1.8 µm, an array pitch of 3.0 µm, and a triangular layout, the open-to-total area ratio is approximately 32.5%. For statistical comparison, we analyzed 40 × 26 µm² regions containing ~135 microholes. Using the estimated $D_{\text{screw}}$ and exposed area, we calculated the probability $P$ of a microhole containing at least one screw dislocation. Applying the complementary probability rule, we obtained the expected number of affected microholes, $N_{\text{screw}} = P \times 135$. This value was then compared with the actual number of taller platelets, $N_{\text{tall}}$, as identified from SEM imaging in the Supplementary Information (Supplementary Figure S2). The resulting data is summarized in Table 1.

Table 1. Full width at half maximum (FWHM) of symmetric (002) reflection rocking curves, calculated screw-type dislocation density ($D_{screw}$), probabilities ($P$) that a micro hole contains at least one screw dislocation, expected number of microholes with screw dislocations ($N_{screw}$) within a 40×26 µm$^2$ area, and the number of tall platelets ($N_{\text{tall}}$) experimentally observed within the same area for the three template types investigated in this study.

| **Template** | **Structures** | **FWHM$_{(002)}$ (arcsec)** | **$D_{\text{screw}}$ (cm$^{-2}$)** | **$P$ %** | **$N_{\text{screw}}$** | **$N_{\text{tall}}$** |
|---|---|---|---|---|---|---|
| Type *A* | GaN/AlN/Sapphire | 83.4 | $1.4\times10^{7}$ | 29.6 | 40 | 35 |
| Type *B* | GaN/Sapphire | 264 | $1.4\times10^{8}$ | 97.0 | 131 | 125 |
| Type *C* | AlN/Sapphire | 17 | $5.8\times10^{5}$ | 1.48 | ~2 | ~0 |

Strikingly, for all template types, the calculated $N_{\text{screw}}$ values match closely with the experimentally observed $N_{\text{tall}}$ counts. This correspondence further confirms that screw dislocations are directly responsible for enhanced vertical growth of *c*-plane facets. The slight underestimation of $N_{\text{tall}}$ relative to $N_{\text{screw}}$ in all cases may stem from statistical deviations or the possibility of some microholes hosting multiple screw dislocations, a factor not accounted for in the model.

These findings highlight the importance of low screw dislocation density substrates for achieving uniform arrays of GaN micro-pyramids or platelets. Additional studies, beyond the scope of this report, reveal that when pyramidal structures are fully formed in a single growth step, distortion is more pronounced on type B templates (rich in tall structures), whereas such distortions are significantly reduced on type *A*/*C* templates with fewer tall platelets.

This can be explained by considering the availability of fast-growing crystallographic orientations, particularly the *c*-plane, during late-stage growth. In arrays with fewer tall pyramids, completed structures cease vertical growth and excess Ga adatoms are readily consumed by surrounding, still-growing lower neighbors. Conversely, in arrays dominated by tall, completed pyramids, the number of actively growing neighbors becomes insufficient to absorb the surplus adatoms. This imbalance leads to a local increase in Ga adlayer thickness and the onset of non-uniform, uncontrolled growth resulting in distorted micro-structures deviating from the right-shaped pyramids.

**Effect of Carrier Gas Composition on GaN Platelet Growth.**

To identify the conditions most suitable to both preserve good *c*-plane surface morphology and low platelet-to-platelet height variability, we varied the growth ambient. The carrier gas ratio of $N_2$ and $H_2$ is known to have a significant effect on the morphology of GaN grown selectively[24]. We investigated this effect by systematically changing the the carrier gas ratios ($H_2$: $N_2$) from pure $H_2$, through ratios of 1:1 and 1:3 to pure $N_2$, while keeping the growth temperature at 890°C (except for the "pure $H_2$" for which it was 855°C). The results are shown in Figure 6, *a*-*d*. The uniformity under different carrier gas ratios was also characterized by AFM, covering an area of 15 × 10 µm² were performed, including approximately 25 platelets, with the coefficient of variation (CV) calculated as the ratio of the height standard deviation to average multiplied by 100%, and plotted in Figure 6, *e*. It is found that the size uniformity is improved significantly in the ambient with nitrogen carrier gas although at the cost of deterioration of *c*-plane surface morphology – an effect similar to the reduction of temperature. With pure hydrogen, the platelets exhibit various heights with the CV of 30.82%. With the increasing nitrogen gas ratios, the CV decreases, with the sample grown in pure nitrogen exhibiting the lowest CV of 4.08%. From the scatter, as shown in Figure 6, *f*, the first two samples have the majority distribution at the mean height range and minority differ significantly from the average, while the pure $N_2$ sample exhibits a more balanced and centered scatter plot, indicating significantly improved height uniformity. In terms of surface morphology, however, it is observed that when the carrier gas is at least 50% $H_2$ (in Figure 6, *b*), the platelet surfaces were much smoother, exhibiting no V-pits on the top, compared to the $N_2$-dominated conditions. Figure 6, *a* is unfortunately an exception from this trend due to the lower growth temperature which causes the same effect; a similar degradation of surface morphology was observed at reduced temperatures in the temperature-dependent growth series discussed above. The improved *c*-plane surface morphology could be attributed to the etching effect of $H_2$ and enhanced adatom mobility in the $H_2$-rich ambient[36,37]. While additional factors such as variations in the V/III ratio, precursor decomposition kinetics, and surface chemical reactions may also influence the growth behavior, a detailed kinetic analysis of these effects is beyond the scope of the present study. By increasing the nitrogen gas percentage from 0, 50, 75 to 100%, the platelets tend to grow more vertically than laterally so that the relative growth rate along *c*-plane is enhanced compared to the slanted crystallographic planes leading to both more pronounced V-pits and better defined semi-facets. It is also worth noting that in the $N_2$-dominated ambient, an unwanted deposition on the $SiN_x$ mask starts to appear (see Figure 6, *c* and *d*), due to the reduced mobility and more favorable spontaneous nucleation conditions[38]. Although it is not that obvious here because of the short growth time, it will have a significant impact if the growth prolongs, i.e. when growing full pyramids.

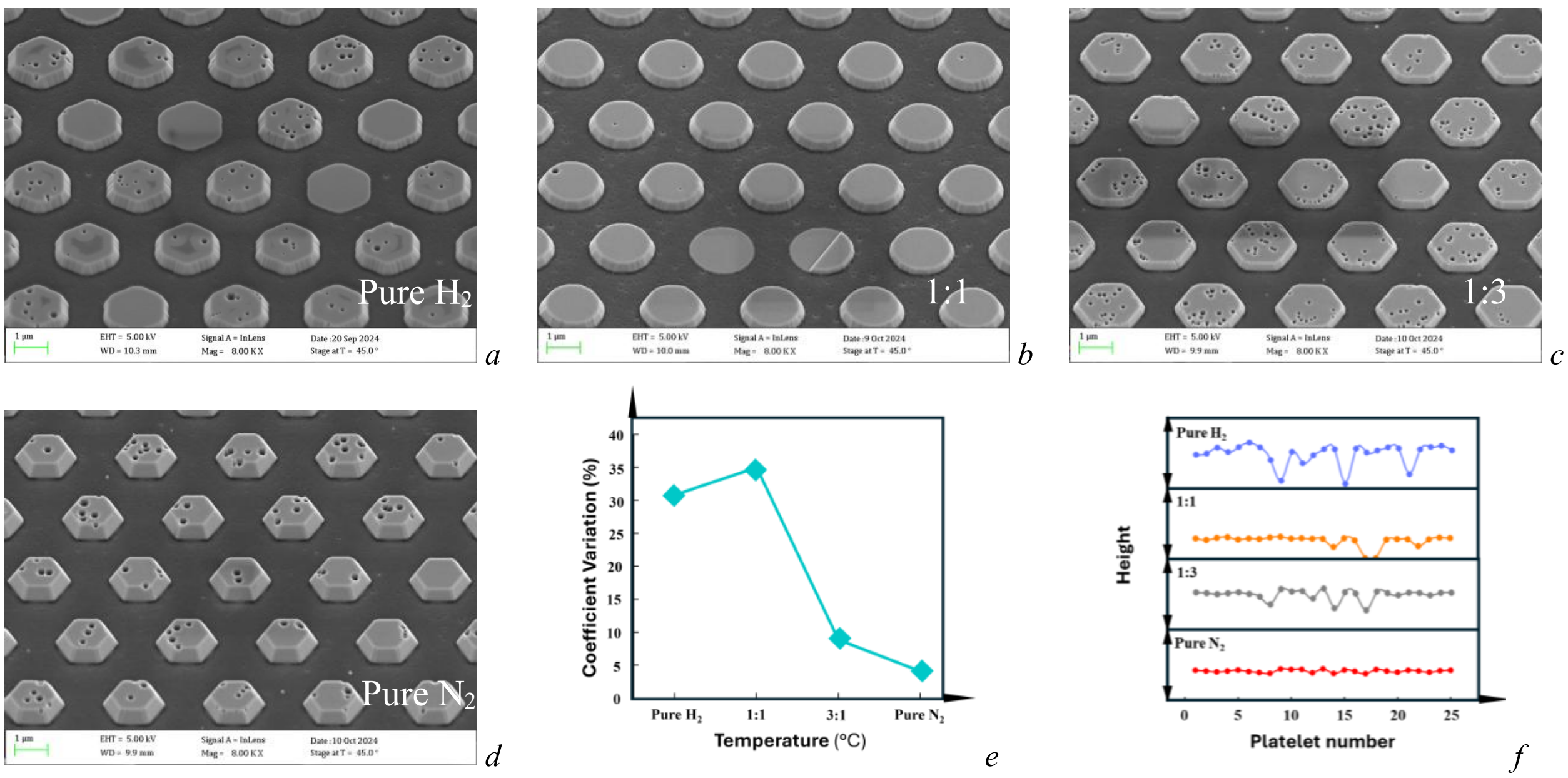

Figure 6. GaN platelets (type *B*) grown under different carrier-gas compositions. SEM images of platelets grown at 855°C in pure $H_2$ (*a*) and at 890°C with $H_2/N_2$ ratios of 1:1 (*b*), 1:3 (*c*), pure $N_2$ (*d*); (*e*) coefficient of variation (CV) of platelet height as a function of carrier-gas composition. , (*f*) normalized by average heights vs platelet number (in arbitrary order) for different carrier gas ratio.

**Thermal annealing of V-pitted micro-platelets in $N_2$+$NH_3$ ambient**

Since finding growth conditions that supporting both smooth *c*-plane surface morphology and an acceptable homogeneity of platelet size on dislocated substrates is challenging, we investigated opportunities to recover *c*-plane morphology by in-situ annealing. Based on previous experience in reshaping GaN nanocolumns which was found to be most efficient at 200 mbar and 950°C in $N_2/NH_3$ ambient[39] and some other reports on curing V-pits at similar temperatures[40], we conducted annealing experiment and managed to successfully recover *c*-plane smoothness of a sample originally grown at a low temperature (Figure 2, *a*), where resulting shape of the micro-platelets can be observed in Figure 7.

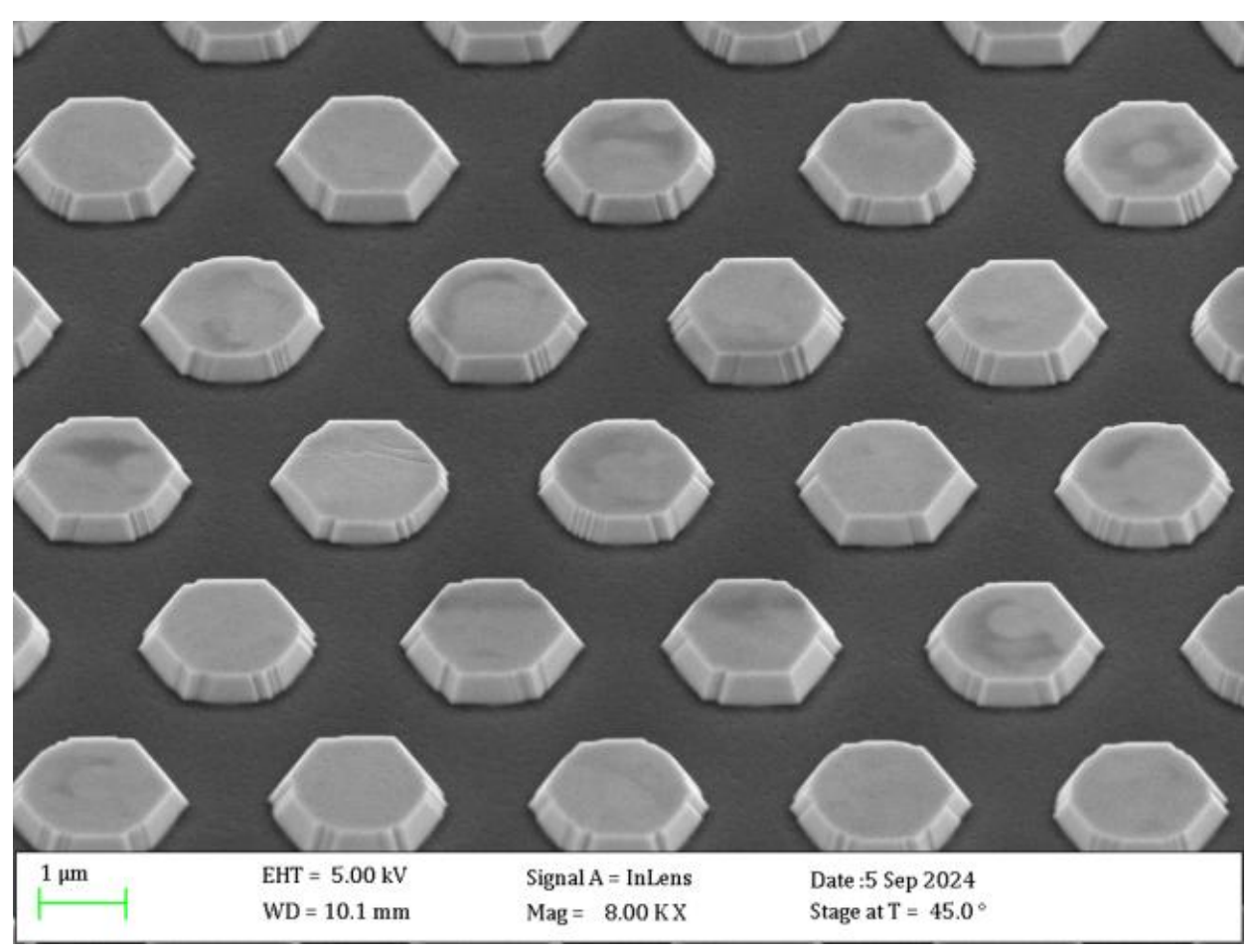

Figure 7. Sample from Figure 2, *a* after annealing in in $N_2$+$NH_3$ ambient at 950°C and 200 mbar for 30 min.

One can see that within 30 minutes of annealing V-pits are completely cured, while the platelets themselves are reshaped towards having more regular hexagonal cross sections. During the high-temperature annealing process, The GaN surface starts to decompose losing Ga atoms from the lattice to a surface Ga adlayer[39]. Each crystallographic orientation of GaN has its own threshold thickness of this adlayer separating conditions of pure decomposition and pure growth.

In the case of annealing, the thickness of the resulting adlayer will be defined by the plane for which this threshold adlayer is the thickest (contributing plane). In this condition, other crystallographic orientations for which this thickness exceeds their equilibrium value, can grow and thus deplete the adlayer locally. Surface adatom diffusion however will equalize the thickness, and the adlayer will be replenished from the decomposition at the contributing crystallographic orientation. In our case, as in our previous work[39] the *c*-plane is the contributing plane while all other orientations are pure "consumers". This allows growth along semi-polar facets while platelets lose a little height. This reduced height and smoother c-plane morphology can be observed comparing Figure 2, *a* and Figure 7.

Based on this finding, we investigated opportunity of growing homogeneous pyramids on imperfect substrates using multi-step growth-then-anneal approach.

**Multi-Step Growth-then-Annealing Approach for Full GaN Pyramid Formation.**

It has been concluded from the previous results that high-quality templates are preferred to obtain uniform GaN platelets or pyramids; however, such templates, for example bulk GaN substrates, are typically associated with high cost. Based on the experiments on GaN platelets, the optimal growth parameters (i.e., pure $H_2$ at 825 °C), which previously yielded the best height uniformity, were therefore employed to develop full pyramidal structures. Two growth approaches were investigated: a direct single-step growth and multi-step growth followed by annealing after each step. For the direct single-step approach, the temperature was maintained at 825 °C for 720 s of continuous growth to form full pyramids, as schematically shown in Figure 8, *a*. The outcome is unfortunately not very impressive as micro-pyramids' tops are almost always deformed by residual V-pits which are not cured even upon pyramid competition (Figure 8, *b* and *c*) significantly affecting the overall pyramid morphology. Additionally, some GaN was deposited on the $SiN_x$ mask, likely due to the excessive growth time (growth after full completion when sample runs out of fast-growing planes) at the low temperature.

To overcome these limitations, a multi-step growth strategy was employed, in which the same total GaN growth time was divided into multiple cycles, each consisting of a shorter growth step followed by annealing in $N_2$ at 950 °C for 600 s. To elucidate the effect of the number of growth–annealing cycles, intermediate cases with two and four cycles were investigated, while keeping the total GaN growth time constant.

As shown in Figure 8, *d-f*, the two-cycle process already leads to a noticeable reduction in both the size and density of V-pits compared to the single-step growth, indicating that the intermediate annealing step partially heals surface defects. When the number of cycles is increased to four (Figure 8, *g-i*), the V-pits become significantly smaller and less frequent, and in some cases are completely absent, demonstrating an enhanced effectiveness of the annealing process. Nevertheless, despite this improvement, the four-cycle process remains insufficient to fully suppress V-pit formation across the entire sample. This suggests that the amount of GaN deposited during each growth step remains too large to be completely removed during the subsequent annealing By further increasing the number of cycles to six (Figure 8, *j-l*), the amount of GaN deposited during each growth step is sufficiently reduced, allowing the subsequent annealing step to effectively eliminate V-pits after each cycle. As a result, nearly perfect GaN pyramids with smooth surfaces and high uniformity in both size and shape are achieved, together with negligible deposition on the SiNx mask. In summary, the multi-step growth–annealing approach prevents the enlargement of V-pits by periodically removing them during intermediate annealing steps, leading to the formation of highly uniform GaN pyramids.

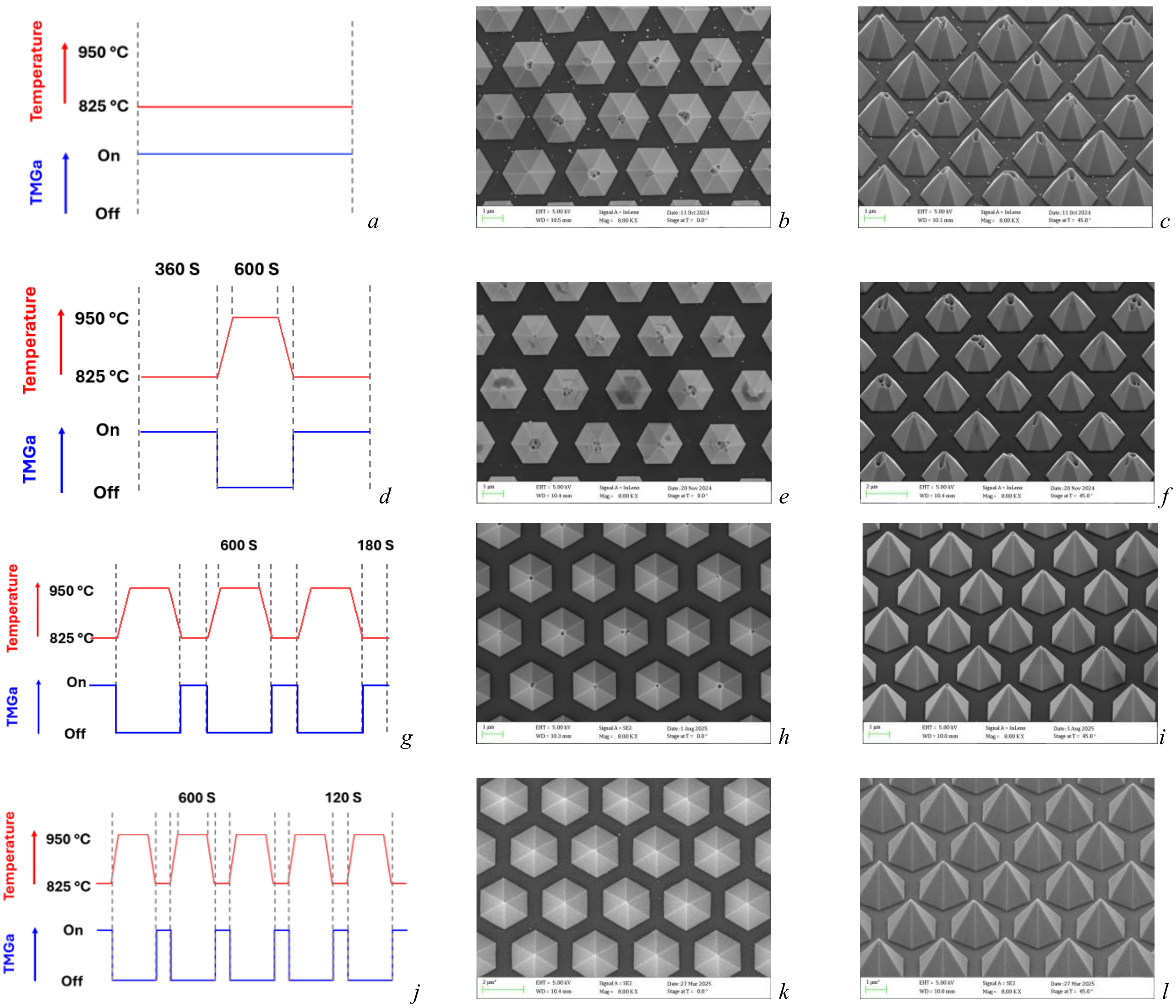

Figure 8. Schematics of the growth temperature and TMGa flow during direct single step (*a*), 2 cycles (d), 4 cycles (g), and 6 cycles multi-step growth-then-annealing (*j*) approaches and top (*b, e, h* and *k*) and bird (*c, f, i* and *l*) view SEMs of the corresponding micro-pyramids.

## CONCLUSIONS

In summary, our study systematically investigated controllable approaches to achieve uniform GaN micro-platelets and pyramids via selective-area growth by addressing critical factors: template dislocation density, growth temperature, carrier gas composition, and multi-step growth-then-annealing strategy. The template selection critically influences morphology uniformity, with low screw-dislocation-density substrates ($D_{screw} \approx 5.8\times10^5$ cm$^{-2}$) enabling superior height uniformity through step-flow dominated growth. Temperature optimization studies revealed that low growth temperature (at 825°C) in $H_2$-rich ambient produced uniform platelets but with a high V-pit density, which can be cured by subsequent annealing at 950°C under $N_2$/ammonia for surface reshaping and improved uniformity. A multi-step growth-then-annealing approach was developed and successfully demonstrated to be able to address the V-pit propagation in pyramidal structures. By dividing the growth into six discrete cycles (120 s growth/600 s annealing at 950°C), we achieved GaN pyramids with excellent hexagonal symmetry and height uniformity, while completely suppressing V-pit evolution and propagation. These studies to achieve uniform GaN platelets and pyramids pave the way for relevant semiconductor devices applications such as micro-LEDs where uniform arrays are essential for performance and yield.

## SUPPORTING INFORMATION

Supporting Information is available free of charge via the ACS Publications website. It includes AFM surface morphology of the three templates and additional SEM images of GaN platelets grown on different templates under identical conditions, with labeled tall and short platelet populations used for statistical comparison. The SEM images of GaN pyramids grown on the patterned templates with various opening diameters and pitches are also shown.

## AUTHOR INFORMATION

### Corresponding Author

**Zhi Li** - *Tyndall National Institute, University College Cork, T12 R5CP Cork, Ireland*;
Orcid link: https://orcid.org/0000-0002-0417-832X; Email: zhi.li@tyndall.ie

### Authors

**Changhao Li** - *Tyndall National Institute, University College Cork, T12 R5CP Cork, Ireland*;
Orcid link: https://orcid.org/0000-0001-9072-5889; Email: changhao.li@tyndall.ie

**Vitaly Z. Zubialevich** - *Tyndall National Institute, University College Cork, T12 R5CP Cork, Ireland*; Orcid link: http://orcid.org/0000-0003-4783-5104; Email: vitaly.zubialevich@tyndall.ie

**Peter J. Parbrook** - *Tyndall National Institute, University College Cork, T12 R5CP Cork, Ireland*; *School of Engineering, University College Cork, Cork, Ireland;* Orcid link: https://orcid.org/0000-0003-3287-512X; Email: peter.parbrook@tyndall.ie

**Brain. Corbett** - *Tyndall National Institute, University College Cork, T12 R5CP Cork, Ireland*; Orcid link: https://orcid.org/0000-0002-9002-8212; Email: brain.corbett@tyndall.ie

### Author Contributions

Z.L. conceived the research project. V.Z and C.L. designed and performed experiments on growth. C.L. conducted fabrication and characterization. All authors discussed the data and wrote the paper.

### Funding Sources

This work is funded by Research Ireland Pathway Program (SFI-IRC_22/PATH-S/10800) and Irish Photonics Integration Centre (IPIC) (SFI-12/RC/2276_P2_IPIC).

## ACKNOWLEDGMENT

All the authors gratefully acknowledge the funding from Science Foundation Ireland (SFI) Pathway Program (SFI-IRC_22/PATH-S/10800) and Irish Photonics Integration Centre (IPIC) EMERGE Program (SFI-12/RC/2276_P2_IPIC). The authors thank Dr. Brendan Sheehan for his support with AFM characterization and Dr. Pietro Pampili for fruitful discussions regarding the growth issues.

# For Table of Contents use Only

## High Uniformity GaN Micro-pyramids and Platelets by Selective Area Growth

Changhao Li[1], Vitaly Z. Zubialevich[1], Peter J. Parbrook[1, 2], Brian Corbett[1], and Zhi Li[1*]

[1] Tyndall National Institute, University College Cork, Lee Maltings, T12 R5CP, Cork, Ireland

[2] School of Engineering, University College Cork, College Road, T12 HW58, Cork, Ireland

*Email: zhi.li@tyndall.ie

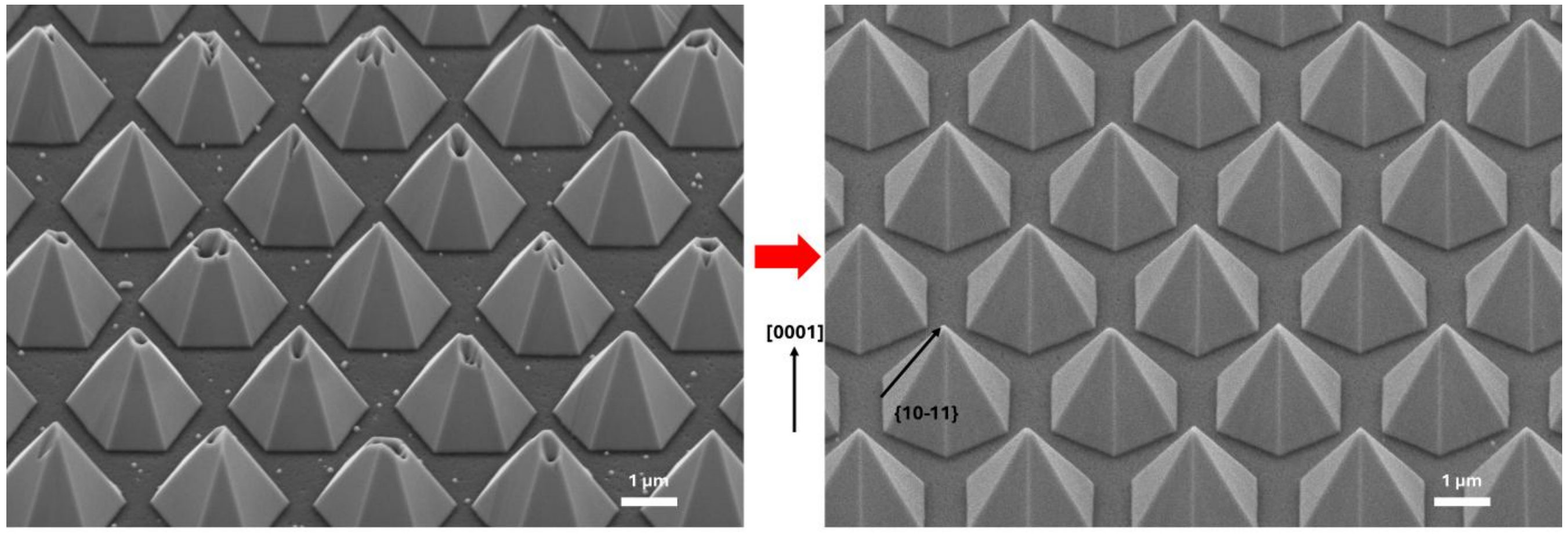

A multi-step growth-then-annealing approach is introduced to achieve highly uniform GaN micro-pyramids by suppressing V-pit formation and propagation. Periodic annealing enables smooth facet reshaping, leading to pyramids with well-defined hexagonal symmetry and excellent height uniformity across arrays.